\documentclass[12pt,a4paper]{article}
\usepackage[utf8]{inputenc}
\usepackage[T1]{fontenc}
\usepackage[english]{babel}
\usepackage{amsmath,amsthm}
\usepackage{amsfonts}
\usepackage{amssymb}
\usepackage{graphicx}
\usepackage{subfig}
\usepackage[left=2cm,right=2cm,top=2cm,bottom=2cm]{geometry}
\usepackage{authblk}
\usepackage{xcolor}

\title{X-ray flares and coronal mass ejections (CMEs) during very quiet solar activity conditions of 2009}

\author[$\star$ $\#$]{Kamsali Nagaraja}
\author[$\star$]{Praveen Kumar Basuvaraj}
\author[$\star$]{S. C. Chakravarty}

\affil[$\star$]{Department of Physics, Bangalore University, Bengaluru 560056 India}
\affil[$\#$]{Corresponding author: \texttt{kamsalinagaraj@bub.ernet.in}}
\date{}

\begin{document}

\maketitle

\begin{abstract}
Solar flares (SFs) are sudden brightening observed over the Sun’s surface which is associated with a large energy release. Flares with burst of X-ray emission are normally followed by a mass ejection of electrons and ions from the solar atmosphere called Coronal Mass Ejections (CMEs). There is an evidence that solar magnetic field can change its configuration through reconnection and release energy, accelerating solar plasma causing SFs and CMEs. This study examines the SFs/CMEs data from SOHO and GOES satellites during the very low solar activity year of 2009 and moderately high solar activity of 2002. The results indicate that certain modifications in the existing mechanisms of generating SFs/CMEs would be necessary for  developing more realistic forecast models affecting the space weather conditions.
\end{abstract}


\section{Introduction}
Solar activity is associated with the occurrence of sunspots on the photosphere with a nominal cycle of $\approx$11 years. The peak sunspot number (SSN) and the period vary from cycle to cycle. The year 2009 witnessed one of the lowest sunspot numbers in recent years. The active periods around the SSN peak years are known for enhanced emission of solar X-ray and energetic charged particle radiations. The solar flares (SF) and coronal mass ejections (CMEs) are highly energetic events producing bursts of these electromagnetic and charged particle radiations, respectively, and involving energies higher than $10^{32}$ ergs, where the event lasts for few minutes to several hours. Model studies suggest that the inverted ‘Y’ or ‘X’ shaped topology of magnetic reconnection in the vertical current sheet produced by the preceding CMEs or SF from below the photosphere as the possible cause of such eruptions. Magnetic reconnection, the process by which magnetic lines of force break and rejoin into a lower-energy configuration, is considered to be the fundamental process by which magnetic energy is converted into plasma kinetic energy. Sun has a large reservoir of magnetic energy, and the energy released by magnetic reconnection has been invoked to explain both large-scale events, such as SF and CMEs [1].

The primary energy release due to reconnection takes place high in the corona that heats up the plasma and accelerates the charged particles. Some particles escape through the open magnetic field lines into the interplanetary space arriving at 1 AU, other particles (electrons and ions) travel back through the newly created magnetic loop towards lower regions of solar atmosphere producing Bremsstrahlung X-rays through Coulomb collisions with ambient plasma. Accelerated ions, by colliding with background particles can excite nuclear reactions and produce $\gamma$ ray emission [2]. Solar flares are often associated with CMEs when huge mass of magnetically confined and accumulated solar plasma erupts from different regions of solar atmosphere. If the CMEs are directed towards earth, it can reach there in 1-3 days, depending on its speed and angular size. These eruptions play a major role in space weather as they cause phenomena such as aurora, geomagnetic  storms, shocks and energetic particle events [3]. The most energetic CMEs occur in close association with powerful flares. Nevertheless large-scale CMEs do occur in the absence of major flares even though these tend to be slower and less energetic.

Solar flare eruptions play a major role in space weather as they cause phenomena such as aurora, geomagnetic storms, shocks and energetic particle events. There is evidence that many energetic CMEs occur in close association with powerful Solar X-ray flares. Nevertheless large numbers of CMEs do occur in the absence of major flares even though these CMEs tend to be slower and hence less energetic.

The main purpose of this study is to examine the combined occurrence and characteristics of SFs and CMEs during a very low solar active period of 2009 determined by the variation of the daily mean sunspot numbers (SSN).

\section{Data Analysis and Methods}
Based on the coronagraphic and X-ray imaging detector observations carried out by the spacecraft SOHO (SOlar and Heliospheric Observatory) and GOES (Geostationary Operational Environmental Satellite), a database has been developed primarily for monitoring/predicting solar energetic proton (SEP) events and its SF/CMEs origin [4] towards a global initiative to space weather studies. This dataset  includes all important parameters related to SEPs/SFs/CMEs covering about 3-solar cycle periods extending from 1984-2013. We have used the SF/CME data sets for the year 2009 and 2002 for detailed investigation of its variability in a very low solar activity as well as a moderately high solar activity year, respectively. These data sets have been supplemented with the international daily SSN data available from the sunspot index and long-term solar observations (SILSO), the World Data Centre for the production, preservation and dissemination of the sunspot numbers. 

\section{Results and Discussion}
The peak values of the solar soft X-ray (SXR) fluxes in the 1-8 \r{A} wavelength range for all solar flare events are collected in each day-bin. If there is more than one flare per day, then the peak flux values for
each flare are kept in the individual day bins as separate events. The flares are categorised based on the peak flux values. Extremely strong flares called X-class of flares are those having peak fluxes greater than $10^{-4}$ W m$^{-2}$.

Figure 1 shows a scatter plot of the peak X-ray fluxes covering all the day-bins of 2009. Also shown in the figure are the daily SSN values. It can be noted that there is a high positive correlation between the occurrence of X-class flares and the values of SSN (which have varied marginally between 0-30 during the whole period). So a small change in solar activity can enable initiation of a solar flare event with an observed SSN threshold value of ~2-3.

Solar flares are often accompanied by the CMEs. To check this aspect, we use the mean speed and angular width parameters to study its variability. The particle radiation of CME may arrive after 2-3 days of the corresponding X-ray flux of the flares. However, no adjustment is made in order to exactly match it with flare times. The categorisation of the CMEs is done based on the mean speed of the event, viz., for speeds below 500 km s$^{-1}$ the CMEs are called C-type (common), for 500-999 km s$^{-1}$, O-type (occasional), for 1000-1999 km s$^{-1}$, R-type (rare) and for 2000-2999 km s$^{-1}$, ER-type (extremely rare).
The number of solar flare events increases when we move from low sunspot to high sunspot years, for example, 2009 and 2002. The trend lines (not plotted here) of the peak X-ray flux and integrated X-ray flux values show a positive correlation between the two parameters for both 2002 and 2009.

\begin{figure}[h]
	\vspace*{0cm}
	\centering
	\makebox[0pt]{%
	\includegraphics[height=7.5cm,width=0.8\paperwidth]{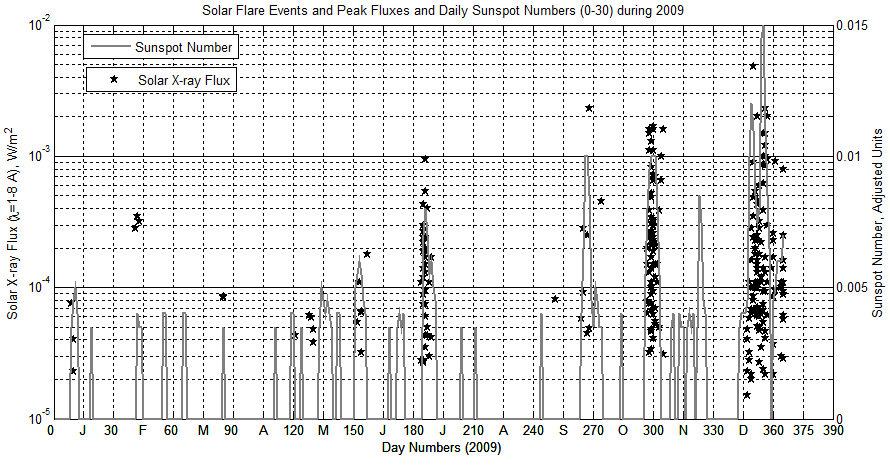}}
	\caption{Peak solar flare time X-ray fluxes (scatter points) and daily SSN (grey lines) plotted covering all the days of 2009. The flux values are shown only if there are solar flare events. Multiple points occur on the same day if there is more than one flare per day.\label{overflow}}
\end{figure}

\begin{figure}[h]
	\vspace*{0cm}
	\centering
	\makebox[0pt]{%
	\includegraphics[height=7.5cm,width=0.8\paperwidth]{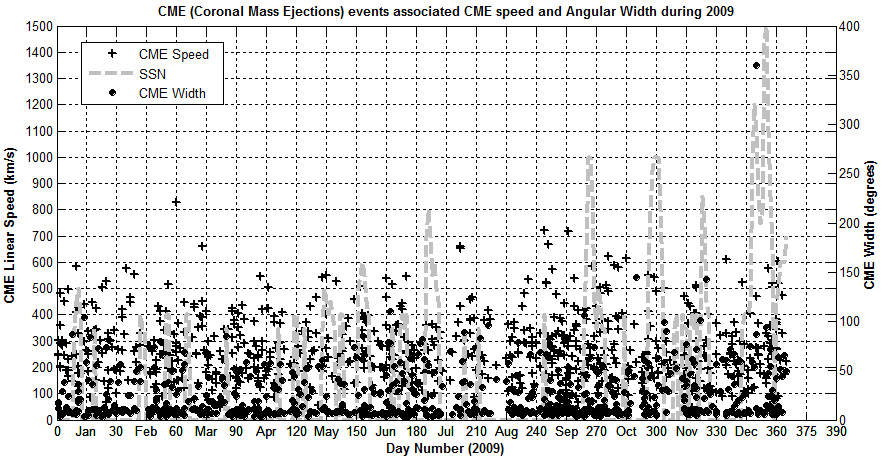}}
	\caption{Scatter plots of CME mean speeds and angular widths for all the events occured during 2009. The daily SSN values are also shown by grey dashed lines. \label{overflow}}
\end{figure}

Figure 2 shows the plots of all the CME events during 2009. Both the mean speed and angular widths of each event are plotted with day numbers. The daily SSNs are also shown in arbitrary units superposed onto the scatter diagram. It can be seen that the majority of CMEs are the common type with only a small percentage of the occasional type but there is no clear correlation with the SSN values. So the origin of the CMEs may not be entirely due to the SF effects.
CMEs events are large in number compared to solar flare but there are a very few events having linear speeds of particles higher than 500 km s$^{-1}$. It is also observed that, there is no significant correlation between the occurrence of high speed CME events and solar activity above SSN > 50.

The relation between SF fluxes and CMES speed cannot be readily discerned from this analysis as that would require data on the location and dynamics of the active region on the sun. But with the CMES speed, it is possible to back calculate the starting time of the event near the solar chromosphere/corona.

\section*{Acknowledgement}
The authors are thankful to the teams of WDC/SILSO for international daily sunspot numbers, NASA/SOHO/GOES for solar flares and CMEs data sets made available in their websites. We acknowledge the utility of these data sets. Authors are also thankful to Indian Space Research Organisation (ISRO), Government of India for financial support.  

\section*{References}
\begin{description}
	
\item [1.] Innes, D. E., Inhester, B., Axford, W. I., Wilheim, K., Nature, 386 (1997) 811.

\item [2.] Krucher, S., Fivian, D., Lin, R. P., Adv Space Res, 35 (2005) 1707.

\item [3.] Priest, E., Magnetohydrodynamics of the Sun, Cambridge Univ Press, United Kingdom (2014).

\item [4.] Papaioannou, A., Sandberg, I., Anastasiadis, A., Kouloumvakos, A., Georgoulis, K., Tziotziou, K., Tsiropoula, G., Jiggens, P., Hilgers, A., J Space Weather Space Clim, 6 (2016).

\end{description}

\end{document}